\documentclass[showpacs,amsmath,amssymb,pre]{revtex4}
\usepackage{bm}
\usepackage{graphicx}

\renewcommand{\vec}[1]{\bm{#1}}
\newcommand{\op}[1]{\mathcal{#1}}
\newcommand{\mat}[1]{\mathsf{#1}}
\newcommand{\ang}{\widehat{\nabla}}
\begin{document}
\title{Spin Dynamics for the Lebwohl-Lasher Model}%
\author{Michael P. Allen}
\affiliation{Department of Physics, University of Warwick, 
  Coventry CV4 7AL, United Kingdom}
\date{\today}%
\begin{abstract}
A spin dynamics algorithm, combining checkerboard updating and a rotation
algorithm based on the local second-rank ordering field, is developed for the
Lebwohl-Lasher model of liquid crystals. The method is shown to conserve energy
well and to generate simulation averages which are consistent with those
obtained by Monte Carlo simulation. However, care must be taken to avoid the
undesirable effects of director rotation, and a method for doing this is proposed.
\end{abstract}
\maketitle

Many physical systems may be represented by the highly idealized model of a set
of interacting classical spin vectors located on a regular, often cubic, lattice
\cite{landau.dp:2000.a}. This paper considers a classical Hamiltonian of
the form
\begin{equation}\label{eqn:Ham}
H = \sum_{\langle j,k\rangle} h(\vec{s}_j\cdot\vec{s}_k) 
= \tfrac{1}{2}\sum_j\sum_{k\in N_j} h(\vec{s}_j\cdot\vec{s}_k)
\end{equation}
where the interaction energy is given by
\begin{equation}\label{eqn:ham}
h(\vec{s}_j\cdot\vec{s}_k) =
 -J 
\left( \tfrac{3}{2}\bigl(\vec{s}_j \cdot
  \vec{s}_k\bigr)^2 - \tfrac{1}{2} \right)
\end{equation}
In eqn~\eqref{eqn:Ham}, the notation $\sum_{\langle j,k\rangle}$ indicates a sum
over all nearest neighbour lattice sites $j$, $k$, considering each neighbour
pair only once. Correspondingly $\sum_{k\in N_j}$ denotes a sum over sites $k$
which constitute the set $N_j$ of nearest neighbours of $j$. Full periodic
boundary conditions are assumed.  The spins $\vec{s}_j$ are three-dimensional
vectors of unit length, and the pair interaction has the form of a simple
function $h$ of the scalar product of the corresponding vectors. The case
$h(\vec{s}_j\cdot\vec{s}_k) = -J \vec{s}_j\cdot\vec{s}_k$ is the well-studied
classical Heisenberg model, which, for $J>0$, exhibits a ferromagnetic ground
state. Our interest lies in the model defined by eqn~\eqref{eqn:ham}, originally
proposed by \citet{lebwohl.pa:1972.a,lebwohl.pa:1973.a} to represent the
ordering in nematic liquid crystals. This model has been extensively studied by
Monte Carlo simulations \cite{landau.dp:2000.a} in the canonical ensemble
\cite{luckhurst.gr:1982.a,fabbri.u:1986.a,zhang.z:1993.a}. At high temperature
$T$ the system is disordered, while for $J>0$ the low-temperature phase has
aligned spins; the definition of an order parameter will be given below. The
phase transition is known to be weakly first order. Brownian or Langevin
dynamics have also been applied to this model
\citep{ding.jd:1996.a,bradac.z:1998.a,bac.cg:2001.a,svetec.m:2004.a}; here the
approach of spin dynamics is considered.

The spin-dynamics equations of motion take the form
\begin{equation}
\dot{\vec{s}}_j =
\frac{\partial H}{\partial\vec{s}_j}
\times \vec{s}_j 
\equiv \vec{\Omega}_j \times \vec{s}_j 
\end{equation}
the dot denotes time differentiation, and $\times$ is the vector product.
The local field $\partial H/\partial\vec{s}_j$ at spin $j$ 
defines an instantaneous angular velocity of precession, $\vec{\Omega}_j$.
This dynamics conserves the individual spin lengths, as may be seen by
considering the time derivative of $|\vec{s}_j|^2$:
\begin{equation*}
\frac{\text{d}}{\text{d}t} |\vec{s}_j|^2 =
2 \dot{\vec{s}}_j \cdot\vec{s}_j =
2 \bigl(\vec{\Omega}_j \times \vec{s}_j\bigr)\cdot\vec{s}_j = 0 \:.
\end{equation*}
The hamiltonian is also conserved:
\begin{equation*}
\dot{H}  =
\sum_j \frac{\partial H}{\partial\vec{s}_j}
\cdot \dot{\vec{s}}_j = 
\sum_j \vec{\Omega}_j
\cdot \bigl(\vec{\Omega}_j \times \vec{s}_j\bigr) = 
0 \:.
\end{equation*}
Finally, for the class of hamiltonians of eqn~\eqref{eqn:Ham}, the total
  magnetization $\vec{S}=\sum_j\vec{s}_j$ is also conserved:
\begin{equation*}
\dot{\vec{S}} = \sum_j \dot{\vec{s}}_j
= \sum_j \frac{\partial H}{\partial\vec{s}_j}
\times \vec{s}_j
= \sum_j \sum_{k\in N_j} h'(\vec{s}_j\cdot\vec{s}_k) 
\vec{s}_k\times \vec{s}_j
= 0 \:.
\end{equation*}
Here $h'$ stands for the derivative of $h$ with respect to its
argument, and the right hand side vanishes because every pair
interaction is included twice, once as $(j,k)$ and once as $(k,j)$,
and these cancel because $\vec{s}_k\times
\vec{s}_j=-\vec{s}_j\times \vec{s}_k$.

An algorithm to simulate the spin dynamics of the Heisenberg model was developed
independently by \citet{frank.j:1997.a} and \citet{krech.m:1998.a}.  The set of
all spins $\vec{s}\equiv\{\vec{s}_j\}$ is subdivided into two sublattices, A and
B, in a checkerboard fashion. All spins on sublattice A interact only with their
nearest neighbours, which are all on sublattice B, and vice versa. Formally, the
generator of infinitesimal rotations of the whole set of spins may be decomposed
into separate, non-commuting, matrices or operators which act on the
corresponding sublattices. A time-reversible approximation to the full rotation
operator acting over a finite timestep may be formally written
\begin{equation}\label{eqn:Rotate}
\vec{s}(\Delta t)
\approx 
\op{B}(\tfrac{1}{2}\Delta t)
\op{A}(\Delta t)
\op{B}(\tfrac{1}{2}\Delta t) \vec{s} +
\mathcal{O}(\Delta t^3)
\end{equation}
The operator $\op{A}(\Delta t)$ represents the rotation of the A-spins in an
external field caused by the neighbouring fixed B-spins, during a period $\Delta
t$; similarly for $\op{B}$.  The algorithm proceeds in an alternating fashion:
first solving the dynamics of the spins on one sublattice, with the other
sublattice spins held fixed; then vice versa, and so on. For the detailed
justification of this algorithm, see \cite{frank.j:1997.a,krech.m:1998.a}.

For the Heisenberg model, this decomposition is especially convenient, since
$\vec{\Omega}_j$ does not depend on $\vec{s}_j $, and the motion of spins on a
sublattice $\dot{\vec{s}}_j = \vec{\Omega}_j \times \vec{s}_j$ with constant
$\vec{\Omega}_j$ may be solved \emph{exactly}. Set
$\Omega_j=|\vec{\Omega}_j|$, and $\hat{\vec{\Omega}}_j=\vec{\Omega}_j/\Omega_j$
in the finite rotation formula \citep{goldstein.h:1980.a} to give
\begin{equation}
\vec{s}_j(\Delta t) = 
\bigl(\hat{\vec{\Omega}}_j\cdot\vec{s}_j\bigr)\hat{\vec{\Omega}}_j
+
\sin (\Omega_j \Delta t) \; \hat{\vec{\Omega}}_j\times\vec{s}_j
+
\cos (\Omega_j \Delta t) \; \left(\vec{s}_j - 
\bigl(\hat{\vec{\Omega}}_j\cdot\vec{s}_j\bigr)\hat{\vec{\Omega}}_j
\right)
\: .
\label{eqn:frot}
\end{equation}
This represents the practical implementation of one of the $\op{A}$ or $\op{B}$
sub-steps in eqn~\eqref{eqn:Rotate}.

For the Lebwohl-Lasher model of eqn~\eqref{eqn:ham}, 
\begin{equation*}
\vec{\Omega}_j = -3J \sum_{k\in N_j} \bigl(\vec{s}_j \cdot
  \vec{s}_k\bigr) \vec{s}_k \:,
\end{equation*}
the motion during one time step is no longer a simple rotation, because although
the neighbouring spins $\vec{s}_k$ are fixed, the moving spin $\vec{s}_j$
appears on the right.  \citet{krech.m:1998.a} propose an iterative approach to
this problem; here a different method is adopted. The equation of motion of each
spin is conveniently written
\begin{equation*}
\dot{\vec{s}}_j =
\vec{\Omega}_j\times \vec{s}_j
= -3J \sum_{k\in N_j} \bigl(\vec{s}_j \cdot
  \vec{s}_k\bigr) \vec{s}_k \times \vec{s}_j
\equiv \vec{s}_j \cdot \mat{F}_j \times \vec{s}_j
\end{equation*}
defining a tensor field at each lattice site due to the
neighbouring spins
\begin{equation*}
\mat{F}_j = -3J \sum_{k\in N_j}
\left(\vec{s}_k\otimes\vec{s}_k - \tfrac{1}{3}\mat{1}\right) \:.
\end{equation*}
A term $\frac{1}{3}\mat{1}$, where $\mat{1}$ is the unit tensor, is subtracted
to make $\mat{F}_j$ traceless: this has no effect on the equations of motion since
$\vec{s}_j \cdot \mat{1} \times \vec{s}_j = \vec{s}_j \times \vec{s}_j = 0$.

The above equation is well known in another context: the torque-free time
evolution of the angular velocity of a rigid body, expressed in the frame of
reference of the inertia tensor of the body itself \cite{goldstein.h:1980.a}.
The symmetric tensor $\mat{F}_j$ plays the role of the inertia, but it arises
from a different mechanism here. A method for integrating this over a timestep
has been proposed \citep{dullweber.a:1997.b}. It is convenient to resolve all
the vectors in the principal axis system of $\mat{F}_j$.  Denote the
eigenvalues, in order $F_{j1}>F_{j2}>F_{j3}$, and the corresponding eigenvectors
$\vec{1}_j,\vec{2}_j,\vec{3}_j$. These are taken to be mutually perpendicular
and of unit length, and all the following vector and matrix expressions are
expressed in this frame.  It should be noted that the algorithm is independent
of the sign ambiguity associated with these eigenvectors. $\mat{F}_j$ becomes
diagonal, and the instantaneous angular velocity takes a simple form:
\begin{equation*}
\mat{F}_j=\begin{pmatrix} F_{j1} & 0 & 0 \\ 0 & F_{j2} & 0 \\ 0 & 0 & F_{j3}
\end{pmatrix}\;,
\qquad\Rightarrow
\qquad
\vec{\Omega}_j=\begin{pmatrix} s_{j1}F_{j1} \\ s_{j2} F_{j2} \\ s_{j3} F_{j3}
\end{pmatrix}
\end{equation*}
The equations of motion $\dot{\vec{s}}_j=\vec{\Omega}_j\times\vec{s}_j$ become
\begin{equation*}
\dot{s}_{j1} = \bigl(F_{j2}-F_{j3}\bigr) s_{j2}s_{j3}\;,\quad
\text{and cyclic permutations.}
\end{equation*}
Consider the component involving $F_{j1}$, generating rotations about the
$\vec{1}_j$ axis (the others are similar):
\begin{equation*}
\dot{s}_{j1} = 0\;, \quad
\dot{s}_{j2} = -F_{j1}s_{j1}\, s_{j3}\;, \quad
\dot{s}_{j3} =  F_{j1}s_{j1}\, s_{j2}
\end{equation*}
During the action of this part of the field, the component $s_{j1}$ remains
constant, and the other two components are rotated about the $\vec{1}_j$
direction at an angular velocity $\Omega_{j1}=F_{j1}s_{j1}$. Suppose the
generator of this infinitesimal rotation, for the A sublattice, is represented
by the operator $\op{A}_1$. This is combined with the generators of rotations
about the other two axes, and approximated over a finite interval in the
following symmetric decomposition \citep{dullweber.a:1997.b}:
\begin{equation}\label{eqn:rotate}
\vec{s}(\Delta t) 
\approx 
\op{A}_3(\tfrac{1}{2}\Delta t)
\op{A}_2(\tfrac{1}{2}\Delta t)
\op{A}_1(\Delta t)
\op{A}_2(\tfrac{1}{2}\Delta t)
\op{A}_3(\tfrac{1}{2}\Delta t)
\vec{s}
\end{equation}
A similar approach is applied to the B-lattice updates.  Each separate rotation
is implemented with the finite rotation formula \eqref{eqn:frot}.  In the above
equation, the rotation $\op{A}_1$ associated with the largest eigenvalue
$F_{j1}$ is centrally placed, and for brevity this is denoted the \textbf{32123}
sequence; an empirical comparison with an alternative \textbf{12321}
sequence is presented below.

Ordering in this model is measured by the symmetric and traceless second-rank
tensor
\begin{equation*}
\mat{Q} = \frac{1}{N} \sum_{j=1}^N \left(\tfrac{3}{2}
\vec{s}_j\otimes\vec{s}_j - \tfrac{1}{2}\mat{1}\right) 
\end{equation*}
Note that this, like the hamiltonian of
eqns~\eqref{eqn:Ham} and \eqref{eqn:ham}, is invariant to all spin flips
$\vec{s}_j\rightarrow-\vec{s}_j$.  The largest eigenvalue of $\mat{Q}$ is
conventionally taken to be the order parameter $Q$, and the corresponding
eigenvector $\vec{n}$ is the director.

Comparison with Monte Carlo simulations is facilitated by evaluating the
``configurational temperature'' introduced by \citet{rugh.hh:1997.a}, and
specifically derived for orientational degrees of freedom by
\citet{chialvo.aa:2001.a}. The relevant expression, in our notation, is
\begin{equation*}
k_\text{B}T = \frac{\sum_j%
\left\langle \bigl| \vec{\ang}_j H \bigr|^2 \right\rangle}%
{\sum_j\left\langle\ang_j^2 H\right\rangle}
= -\frac{1}{12}\frac{\sum_j %
\left\langle \bigl|\dot{\vec{s}}_j\bigr|^2  \right\rangle}%
{\left\langle H\right\rangle}
\end{equation*}
where $\vec{\ang}_j$ is the angular gradient, and $\ang_j^2$ the angular
laplacian, with respect to the orientation of spin $j$. The second expression
above is specific to the Lebwohl-Lasher potential, and is easily obtained from
eqn~\eqref{eqn:ham}. In the results reported below, Boltzmann's constant
$k_\text{B}$, and the strength parameter $J$, are chosen to be unity.

\begin{figure}[htp]
\includegraphics[width=0.35\textwidth,clip]{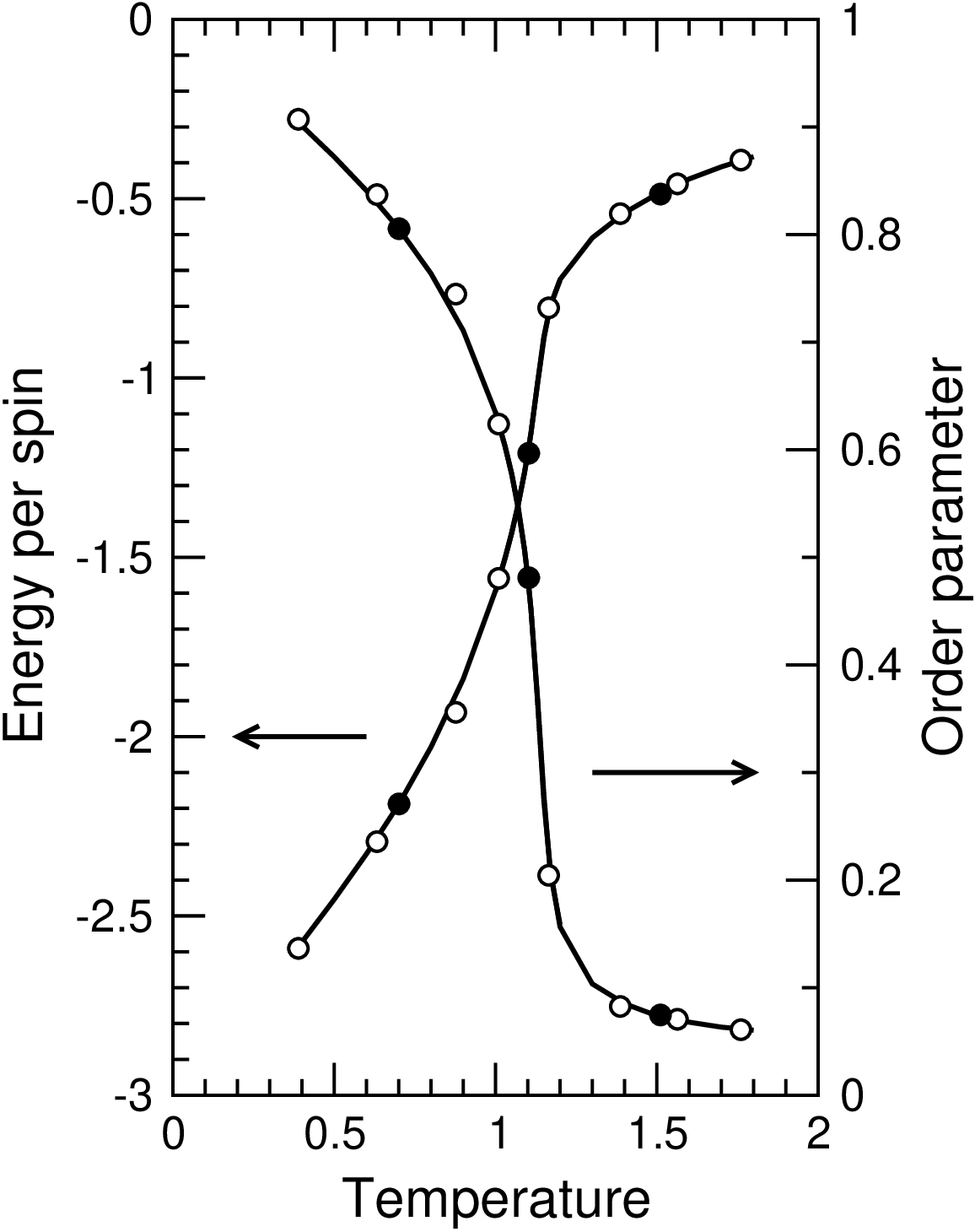}
\caption{\label{fig:1}%
Energy per spin, and nematic order parameter, as functions of
temperature. Lines: Monte Carlo simulations. Circles: spin dynamics. Statistical
errors are smaller than the plotting symbols. Filled symbols indicate state
points studied in more detail below.}
\end{figure}

In Fig.~\ref{fig:1} simulation averages generated by the spin dynamics algorithm
are compared with those produced by conventional Monte Carlo. A system size of
$8\times8\times8$ spins is employed, which is sufficient to show interesting
behaviour in the phase transition region, while being economical.
It is worth emphasizing that the aim is not to properly characterize the
transition, for which much larger systems are required
\citep{luckhurst.gr:1982.a,fabbri.u:1986.a,zhang.z:1993.a}. Spin dynamics runs
of 20000 steps, each of length $\Delta t=0.01$, starting from equilibrated Monte
Carlo configurations, were carried out. Good agreement, within the statistical
errors, is obtained with the Monte Carlo curves. This suggests that the sampling
of the constant-energy hypersurface by spin dynamics generates satisfactory
averages (but see below). Three state points, at $T\approx 1.5, 1.1, 0.7$, with
order parameters $Q\approx 0.0, 0.5, 0.8$ representative of the disordered,
transitional, and ordered states, respectively, were selected for further
illustration.

\begin{figure}[htp]
\includegraphics[width=0.35\textwidth,clip]{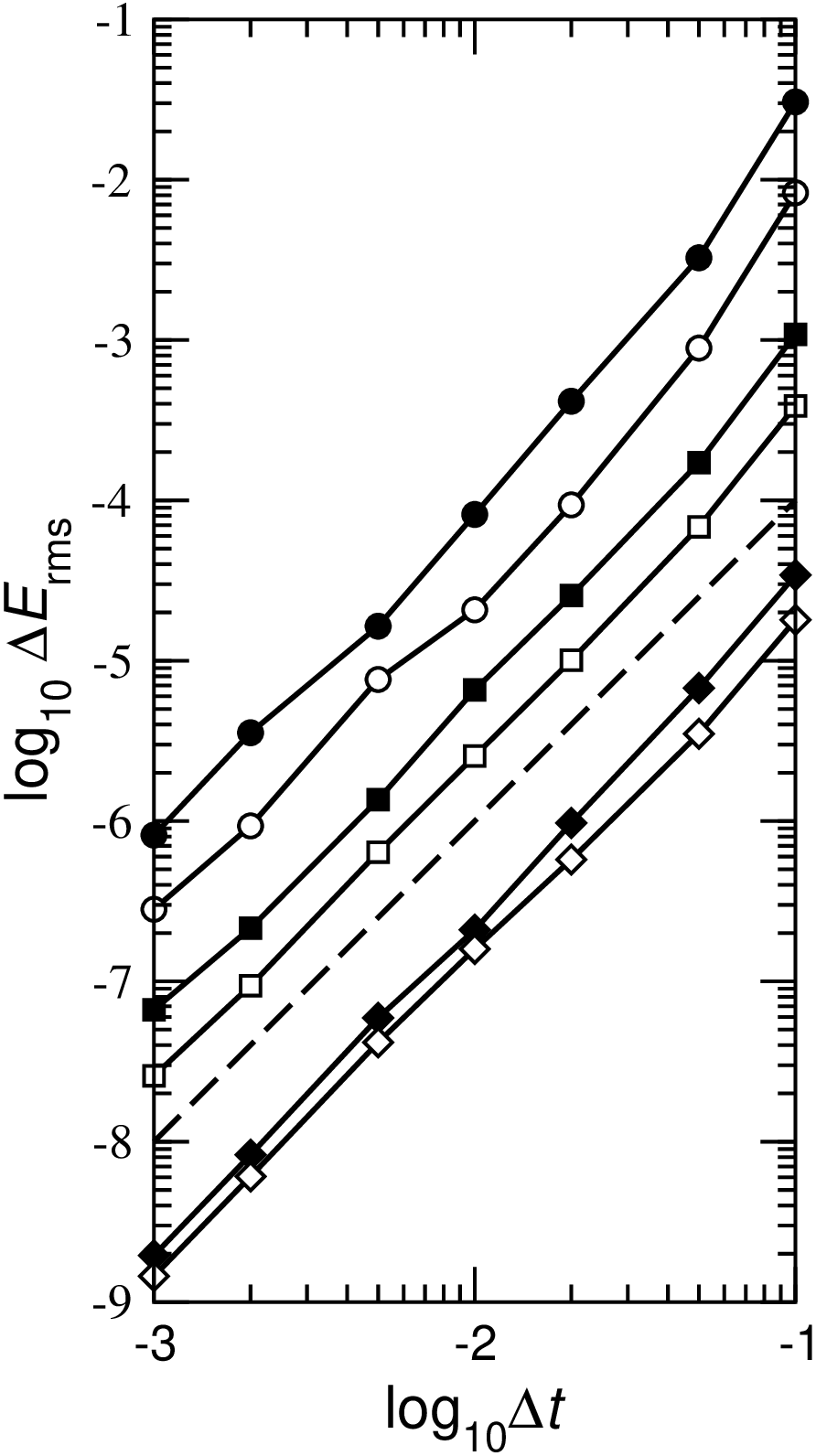}
\caption{\label{fig:2}%
  Root-mean-squared energy fluctuations, plotted against timestep, on
  logarithmic scales. Circles: $T\approx0.7$, $Q\approx0.8$. Squares:
  $T\approx1.1$, $Q\approx 0.5$ (displaced down by a factor of 10 for clarity).
  Diamonds: $T\approx1.5$, $Q\approx 0$ (displaced down by a factor of 100 for
  clarity). Open symbols correspond to the rotation sequence \textbf{32123};
  filled symbols to the sequence \textbf{12321} (see text). The dashed line has
  gradient $2$ for reference.}
\end{figure}

The accuracy of the algorithm, as measured by the root-mean-squared fluctuation
of the energy 
\begin{equation*}
\Delta E_\text{rms}=\sqrt{\langle E^2\rangle-\langle
  E\rangle^2}
\end{equation*}
is presented in Fig.~\ref{fig:2}: $\Delta E_\text{rms}\propto\Delta t^2$ over a
wide range of $\Delta t$. For the largest timesteps there is some drift in the
energy (which is removed in the calculation of $\Delta E_\text{rms}$).
Within each sublattice rotation step,
there are several choices for the sequence of axes about which to carry out the
rotations. The figure shows that the sequence \textbf{12321} (in which the axis
corresponding to the \emph{smallest} eigenvalue of $\mat{F}_j$ is placed
centrally) is worse than the sequence \textbf{32123} of eqn~\eqref{eqn:rotate} (in
which the \emph{largest} eigenvalue is central) by a factor 2--3. Also the
low-temperature ordered state point exhibits worse energy conservation than the
high-temperature, disordered state point, reflecting the larger torques
resulting from the local field.

\begin{figure}[htp]
\includegraphics[width=0.4\textwidth,clip]{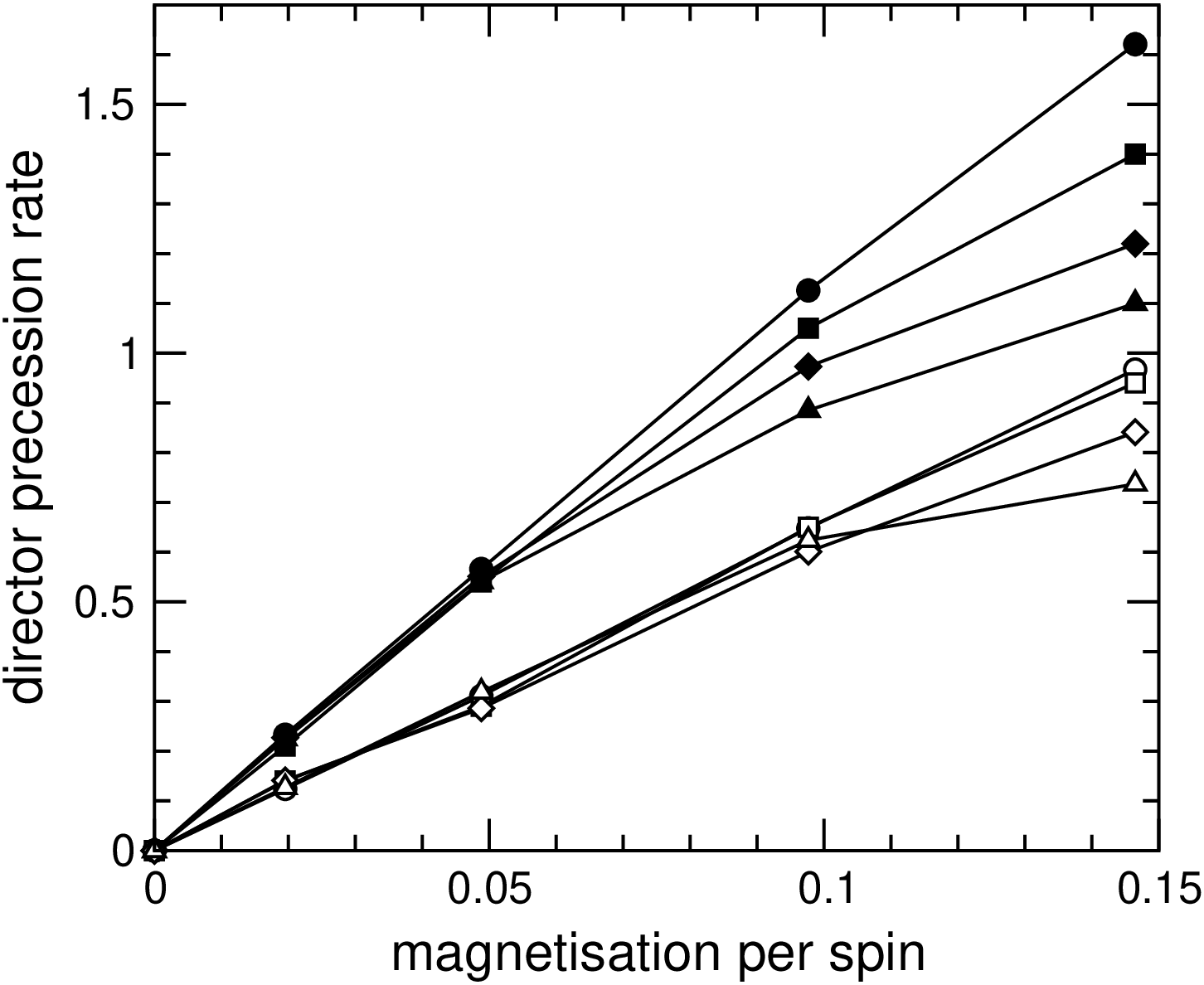}
\caption{\label{fig:3}%
Director rotation as a function of net magnetization per spin. Filled symbols:
$T\approx 0.7$, $Q\approx 0.8$. Open symbols: $T\approx 1.1$,
$Q\approx 0.5$. The director is inclined with respect to the
magnetization vector by $30^\circ$ (circles), $45^\circ$ (squares), $60^\circ$
(diamonds) and $90^\circ$ (triangles).}
\end{figure}

There are some caveats associated with spin dynamics applied to this model.
Firstly, as noted, the total magnetization is conserved exactly by the dynamics,
and asymptotically by the algorithm as $\Delta t\rightarrow 0$. This is an
artificial, physically meaningless, conservation law for the present model.  The
hamiltonian, and the order parameter defined above, are invariant under all spin
flips $\vec{s}_j\rightarrow-\vec{s}_j$, reflecting the symmetry of the nematic
phase.  However, the dynamics is not. If a spin is flipped, it will begin to
rotate in the physically opposite direction. Macroscopic consequences come from
this: in the ordered phase, the director rotates systematically about the fixed
overall magnetization vector, if $\vec{S}=\sum_j\vec{s}_j$ is nonzero.  This
regular precession is superimposed on a general tendency of the director to
become aligned with the magnetization vector: this happens slowly in the ordered
phase, and more rapidly in the transition region.  The rate of precession is
shown in Fig.~\ref{fig:3}, for both the state points, and for a range of tilt
angles of the director relative to the magnetisation, at a range of net
magnetisations obtained simply by flipping spins in an equilibrated Monte Carlo
configuration.
The precession rate is essentially independent of the tilt angle, and is
proportional to both the order parameter $Q$, and to the net magnetization per
spin, as would be expected by considering the interaction between a typical spin
and the local field.

This effect is a potential pitfall in simulations of these systems by spin
dynamics. In a typical configuration of $N$ spins, the net magnetization per
spin will statistically be of order $1/\sqrt{N}$: this will produce a long-lived
slow rotation of the director which, if uncorrected, will dominate measured
dynamical properties. If the magnetization is, for any reason, substantial, the
fast director precession distorts the measured simulation averages, such as
configurational temperature. However, the solution to this problem seems
relatively straighforward. Since the statistical properties of the
Lebwohl-Lasher model are invariant to spin flips, it should be satisfactory to
fix the system magnetization at the minimum possible magnitude, by flipping
spins, at the start of a simulation run. In addition, once the director is
aligned with the magnetization, the effects seem to be minimal.

In this paper, an algorithm for simulating the Lebwohl-Lasher model by spin
dynamics has been presented. It remains to be seen whether this approach will
lead to the determination of interesting ``dynamical'' properties of this model,
and related models, and possibly to accelerated sampling of the transition
region.
\begin{acknowledgments}
The simulations reported here were performed on the computing facilities of the
Centre for Scientific Computing, University of Warwick.
\end{acknowledgments}
\bibliography{journals,main}
\end{document}